\documentclass[aps,showpacs,11pt]{revtex4}
\usepackage{dcolumn}
\usepackage{graphicx}
\usepackage{amsmath}
\usepackage{amsfonts}
\usepackage{amssymb}
\usepackage{psfrag}
\usepackage{wrapfig}
\usepackage{subfigure}
\usepackage{makeidx}
\usepackage{bm}
\usepackage{epsf}
\usepackage{hyperref}
\usepackage{color}
\usepackage{multirow,dcolumn}
\usepackage{graphicx}

\begin{document}

\title{Holographic Thermalization in Charged Dilaton Anti-de Sitter Spacetime}

\author{Shao-Jun Zhang}
\email{sjzhang84@hotmail.com}
\author{E. Abdalla}
\email{eabdalla@usp.br}
\affiliation{Instituto de
F$\acute{i}$sica, Universidade de S$\tilde{a}$o Paulo, C.P. 66318,
05315-970, S$\tilde{a}$o Paulo, SP, Brazil}
\date{\today}

\begin{abstract}
 We study  holographic thermalization in spacetimes with a chemical potential and a non-trivial dilaton field. Three non-local observables are used to probe the whole process and investigate the effect of the ratio of the chemical potential over temperature $\chi$ and the dilaton-Maxwell coupling constant $\alpha$. It is found that the saturation time is not always a monotonically increasing function of $\chi$, the situation depends on $\alpha$. When $0 \leq\alpha \leq 1$, larger $\chi$ yields longer saturation time, while for $\alpha>1$, the situation becomes more complex.  More interesting, we found that although $\alpha$ indeed has influence on the whole thermalization process, it nearly does not affect the saturation time, which indicates the universality of the saturation time for the dual one-parameter field theories.

\end{abstract}


\maketitle

\section{Introduction}

In the past two decades, one of the most important discoveries in modern theoretical physics is the AdS/CFT correspondence~\cite{Maldacena:1997re,Gubser:1998bc,Witten:1998qj}, relating strongly coupled field theories to weakly coupled gravity. It provides  a powerful tool to study strongly coupled field systems where traditional methods encounter great challenges or even break down. Fruitful results have been obtained by applying this duality to various areas of modern physics, such as condensed matter physics~\cite{Hartnoll:2009sz,Herzog:2009xv,McGreevy:2009xe} and  QCD~\cite{Mateos:2007ay,Gubser:2009md,CasalderreySolana:2011us}.

One noticeable application of this duality is to study non-equilibrium physics, in particular in the so-called "holographic thermalization" topic~\cite{Danielsson:1999fa,Janik:2006gp,Chesler:2009cy,Garfinkle:2011hm,Garfinkle:2011tc,
Bhattacharyya:2009uu,Lin:2008rw,Balasubramanian:2010ce,Balasubramanian:2011ur,
AbajoArrastia:2010yt,Aparicio:2011zy,
Albash:2010mv,Keranen:2011xs,Galante:2012pv,Caceres:2012em,Hubeny:2013hz,Li:2013sia,
Li:2013cja,Stricker:2013lma,
Baron:2013cya,Aref'eva:2013wma,Zeng:2013mca,Zeng:2013fsa,Fischler:2013fba,Alishahiha:2014cwa,Fonda:2014ula,Fischler:2014tka,
Pedraza:2014moa,Fischetti:2014hxa,Alishahiha:2014jxa,
Zeng:2014xpa,Zeng:2014xza,Giordano:2014kya,Caceres:2014pda,Camilo:2014,Zhang:2014cga} which is an atempt to describe the thermalization process of the quark-gluon plasma (QGP) produced
in heavy ion collisions at the Relativistic Heavy Ion Collider (RHIC)~\cite{Gelis:2011xw,Iancu:2012xa,Muller:2011ra,CasalderreySolana:2011us}. Although the near-equilibrium physics of the QGP can be understood  using a hydrodynamic description~\cite{Policastro:2001yc,Policastro:2002se,Son:2007vk,Bhattacharyya:2008jc}, the whole process to reach thermal equilibrium is yet not known. Observed data from RHIC show that the time scale needed for QGP to reach thermal equilibrium is considerably shorter than expected based on perturbative methods~\cite{Baier:2000sb,Mueller:2005un}. This indicates that the thermalization process of QGP is strongly coupled, and thus motivates us to use  AdS/CFT relation.

According to AdS/CFT, it is proposed that this thermalization process is dual to the process of a black hole formation in the bulk via gravitational collapse of a scalar field in AdS space~\cite{Danielsson:1999fa,Janik:2006gp,Chesler:2009cy,Garfinkle:2011hm,Garfinkle:2011tc}, or to a collapse of a thin-shell matter described by an AdS-Vaidya metric~\cite{Bhattacharyya:2009uu,Lin:2008rw,Balasubramanian:2010ce,Balasubramanian:2011ur}. To probe the detailed process of the thermalization, local-observables in the boundary field theory, such as energy-momentum tensor and its derivatives, are not sufficient. In refs.~\cite{Balasubramanian:2010ce,Balasubramanian:2011ur}, in the model of the collapsing of a thin-shell,  three non-local observables have been proposed to probe the process, the two-point function, the Wilson loop and the entanglement entropy. According to the AdS/CFT relation, in the semiclassical limit these three non-local observables in the boundary field theory can be evaluated by calculating certain geometric quantities in the bulk geometry. By studying the time evolution of these three non-local observables, and under Einstein gravity frame, the thermalization process has thus been studied in detail and some interesting results are found. First, the saturation time which is needed to reach thermal equilibrium depends on the geometric size of the probe in the boundary field theory. Probes with smaller size thermalize faster indicating a "top-down" thermalization mechanism, which is very much in contrast with the standard "bottom-up" paradigm~\cite{Baier:2000sb} from the use of perturbative methods. Moreover, all probes used show a delay in the onset of the thermalization, an apparent non-analyticity in the end of thermalization, the transition to full thermal equilibrium is instantaneous and these features are independent of dimensionality. Finally, for homogeneous initial conditions the entanglement entropy thermalizes slowest, and thus sets a timescale for equilibration that saturates a causality bound.

Later, this study is extended to include the effect of chemical potential in the boundary field theory~\cite{Galante:2012pv,Caceres:2012em} (see also \cite{Giordano:2014kya}), which is usually the case in real heavy ion collision processes. In the gravity side, this is dual to introducing a gauge field and considering charged background geometries. It is found that the time evolution of the three non-local observables exhibit similar behavior as in chargeless case. The arise of the chemical potential affects the saturation time: the larger the chemical potential, the longer the saturation time. Also the study has been extended to other situations by considering higher curvature corrections~\cite{Li:2013cja,Zeng:2013mca,Zeng:2013fsa}, non-linear electromagnetic effect~\cite{Camilo:2014}, angular momentum~\cite{Aref'eva:2013wma}, noncommutative~\cite{Zeng:2014xpa} and hyperscaling violating geometries~\cite{Alishahiha:2014cwa,Fonda:2014ula}. Further, it is extended to study holographic thermalization of field theories living in curved spacetimes~\cite{Fischler:2013fba,Fischler:2014tka,Fischetti:2014hxa,Zhang:2014cga}.

On the other hand, it is believed that Einstein gravity should be considered as an effective description of the underlying quantum gravity theory (string theory for example) at low energy. Thus, it is interesting to study the effect of stringy corrections on the thermalization process. As we have mentioned above, some work has been done in this directions by considering higher curvature corrections~\cite{Li:2013cja,Zeng:2013mca,Zeng:2013fsa,Zhang:2014cga}. It is found that the existence of higher curvature corrections considerably affects the thermalization process. For example, by including the effect of Gauss-Bonnet terms, the authors in refs.~\cite{Li:2013cja,Zeng:2013mca,Zeng:2013fsa,Zhang:2014cga} found that a larger Gauss-Bonnet coupling constant will make the saturation shorter, thus leaving the imprint of this stringy effect. In this paper, we would like to consider another kind of stringy effect, the emergence of a dilaton field. The action we consider is the well-known Einstein-Maxwell-dilaton action which is a low-energy effective action of string theory~\cite{Horne:1992zy}, where the dilaton couples with the Maxwell field in a non-trivial way.

To use the AdS/CFT relation and  study the effect of the dilaton field, we should first find asymptotically AdS solutions with non-trivial dilaton configuration under this frame. It is  a non-trivial task. Amounts of work have been done to explore various solutions of this gravity theory. Non-trivial coupling between the dilaton and the gauge field yields interesting effects on the available spacetimes~\cite{Gibbons:1987ps,Koikawa:1986qu,Brill:1991qe,Garfinkle:1990qj,Gregory:1992kr,
Horowitz:1991cd,Chan:1995fr,Clement:2002mb,Sheykhi:2006ea,Sheykhi:2007gw,Hendi:2008ng}. However, all of these black hole solutions with non-trivial scalar configuration are not aysmptotically AdS. In~\cite{Poletti:1994ff,Poletti:1994ww,Mignemi:1991wa}, the authors even argued that, with only one Liouville-type dilaton potential besides the cosmological constant, there are no dilaton-dS or AdS black hole solutions. After lots of efforts, an aysmptotically AdS black hole solution was found in ref.~\cite{Gao:2004tv,Gao:2005xv}. By choosing an appropriate combination of three Liouville-type dilaton potentials, the authors successfully obtained the static dilaton black hole solutions which are asymptotically (A)dS in four and higher dimensions. This solution has some interesting properties~\cite{Sheykhi:2008rk,Sheykhi:2009pf,Hendi:2010gq,Ong:2012yf} and has shown potential applications in cosmology~\cite{Gao:2006fu}. We will work with this charged dilaton AdS black hole, and see if and how the presence of the non-trivial dilaton field affects the thermalization process.

The paper is organized as follows. In the next section, we briefly introduce the charged dilaton AdS black hole and then generalize it to obtain its Vaidya-like version. In Sec. III, by applying the three non-local observables, we study the detailed process of thermalization to see the influence of the chemical potential and the dilaton field. The last section is devoted to summary and discussions.

\section{Bulk Spacetimes}

In this section, we  briefly introduce the charged dilation black hole in AdS space, and obtain its Vaidya-like version by introducing external matter source.

\subsection{Charged Dilaton Black Hole in Anti-de Sitter Space}

We consider the action of $(n+1)$-dimensional Einstein-Maxwell-dilaton gravity~\cite{Horne:1992zy,Garfinkle:1990qj,Gao:2004tv,Gao:2005xv}
\begin{eqnarray}\label{action}
S=-\frac{1}{16 \pi} \int d^{n+1}x \sqrt{-g} \left(R-\frac{4}{n-1} (\nabla \Phi)^2 - V(\Phi) - e^{-4 \alpha \Phi/(n-1)} F_{\mu\nu} F^{\mu\nu}\right),
\end{eqnarray}
where $\alpha$ is the coupling constant between the dilaton and the Maxwell field. When $\alpha=0$ this action reduces to the usual Einstein-Maxwell-scalar theory, while for $\alpha=1$ the last term in the bracket represents the well-known dilaton-Maxwell coupling that appears in the low energy string action in Einstein's frame~\cite{Garfinkle:1990qj}. The dilaton potential is chosen to take the  form~\cite{Gao:2004tv,Gao:2005xv}
\begin{eqnarray}\label{potential}
V(\Phi) = &&\frac{2 \Lambda}{n(n-2+\alpha^2)^2} \bigg\{-\alpha^2 \left[(n+1)^2-(n+1) \alpha^2 -6(n+1) +\alpha^2 +9\right] e^{-4(n-2)\Phi/[(n-1)\alpha]}\nonumber\\
&&+(n-2)^2 (n-\alpha^2) e^{4\alpha \Phi/(n-1)} + 4 \alpha^2 (n-1) (n-2) e^{-2(n-2-\alpha^2)\Phi/[(n-1)\alpha]}\bigg\},
\end{eqnarray}
where $\Lambda = -\frac{n (n-1)}{2 l^2}$ is the cosmological constant and $l$  the AdS radius. This type of dilaton potential can appear in the compactification of a higher-dimensional theory, including various supergravity models in four dimensions~\cite{Radu:2004xp,Giddings:2003zw}. From the above action, we can see that reversing the sign of $\alpha$ is equivalent to reverse the sign of the dilaton. Thus, without loss of generality, we can restrict $\alpha$ to be non-negative, $\alpha \geq 0$.

The corresponding  equations of motion are
\begin{eqnarray}\label{EoM}
R_{\mu\nu}-\frac{1}{2} g_{\mu\nu} R- \frac{4}{n-1} \left(\partial_\mu \Phi \partial_\nu \Phi-\frac{1}{2} g_{\mu\nu} \left(\nabla \Phi\right)^2-\frac{n-1}{8} g_{\mu\nu} V(\phi)\right)\nonumber\\
-2 e^{-4\alpha \Phi/(n-1)} \left(F_{\mu\rho} F_{\nu}^{~\rho}-\frac{1}{4} g_{\mu\nu} F_{\rho\sigma} F^{\rho\sigma}\right)&=&0,\nonumber\\
\nabla^2 \Phi - \frac{n-1}{8} \frac{\partial V}{\partial \Phi} + \frac{\alpha}{2} e^{-4\alpha \Phi/(n-1)} F_{\rho\sigma} F^{\rho\sigma}&=&0,\nonumber\\
\nabla_\mu \left(e^{-4\alpha \Phi/(n-1)} F^{\mu\nu}\right)&=&0.
\end{eqnarray}
They admit a charged dilaton black brane solution~\cite{Gao:2004tv,Gao:2005xv}
\begin{eqnarray}\label{orginalmetric}
ds^2 &=& -N(\rho) f(\rho) dt^2 +\frac{d\rho^2}{f(\rho)}+ \rho^2 g(\rho) d\vec{x}_{n-1}^2,\nonumber\\
N(\rho) &=& \Upsilon^{-\gamma (n-3)},\qquad f(\rho) =\frac{\rho^2}{\l^2} \Upsilon^{(n-2)\gamma} -\frac{m}{\rho^{n-2}} \Upsilon^{1-\gamma},\nonumber\\
g(\rho) &=& \Upsilon^\gamma, \qquad \qquad \quad~~~ \Upsilon = 1-\left(\frac{b}{\rho}\right)^{n-2},
\end{eqnarray}
where the constant $\gamma = \frac{2\alpha^2}{(n-2) (n-2 +\alpha^2)}$. The dilaton and the electromagnetic fields take the form
\begin{eqnarray}\label{orginalphi}
\Phi(r) = \frac{n-1}{4} \sqrt{\gamma (2+ 2\gamma-n\gamma)} \ln \Upsilon,\qquad F_{t\rho} =-\sqrt{N(\rho)} \frac{e^{4\alpha \Phi/(n-1)}}{g(\rho)^{(n-1)/2} \rho^{n-1}} q\quad .
\end{eqnarray}
The remaining  components of $F_{\mu\nu}$ vanish. The parameters $m$ and $q$ are related to the physical mass $M$ and charge $Q$ of the black brane,
\begin{eqnarray}\label{masscharge}
M = \frac{\Omega_{n-1}}{16 \pi} (n-1) m,\qquad Q = \frac{\Omega_{n-1}}{ 4\pi} q,
\end{eqnarray}
where $\Omega_{n-1}$ is the volume of $d \vec{x}_{n-1}^2$. Hereinafter, without risk of confusion, we simply call the parameters $m$ and $q$ the mass and charge parameter respectively. The constant $b$ is related to the mass and the charge parameters through the relation
\begin{eqnarray} \label{chargeparameter}
q^2 = \frac{(n-1) (n-2)^2}{2(n-2+\alpha^2)} b^{n-2} m.
\end{eqnarray}
We should emphasize here that the solution contains three free parameters: $(\alpha, m, q)$, or equivalently $(\alpha, m, b)$.

When $\alpha=0$, the solution reduces to the well-known Reissner-Nordstr\"{o}m-AdS black brane. The holographic thermalization in this case has been thoroughly discussed in refs.~\cite{Galante:2012pv,Caceres:2012em}. Thus, in the following we will mainly focus on the case with $\alpha > 0$. In this case, we note that the solution is not real for $0<\rho<b$. We should exclude this region from spacetime, what can be achieved by defining a new radial coordinate $r$ as~\cite{Sheykhi:2009pf,Hendi:2010gq}
\begin{eqnarray}
r^2 = \rho^2 -b^2.
\end{eqnarray}
The metric becomes
\begin{eqnarray}\label{metric}
ds^2 &=& -N(r) f(r) dt^2 + \frac{r^2 dr^2}{(r^2+b^2)f(r)} +(r^2+b^2) g(r) d\vec{x}_{n-1}^2,\nonumber\\
N(r) &=& \Gamma^{-(n-3)\gamma},\qquad f(r) = \frac{r^2 +b^2}{\l^2} \Gamma^{(n-2)\gamma} -\frac{m}{(r^2+b^2)^{(n-2)/2}} \Gamma^{1-\gamma},\nonumber\\
g (r) &=& \Gamma^{\gamma},\qquad\qquad\quad~~~ \Gamma = 1-\left(\frac{b}{\sqrt{r^2+b^2}}\right)^{n-2},
\end{eqnarray}
with the coordinate $r$ now valued in the range $0\leq r  < \infty$. The dilaton and the electromagnetic field become
\begin{eqnarray}\label{phi_F}
\Phi(r) &=& \frac{n-1}{4} \sqrt{\gamma (2+ 2\gamma-n\gamma)} \ln \Gamma,\qquad F_{tr} = -\frac{\sqrt{N(r)} r e^{4\alpha \Phi/(n-1)}}{g(r)^{(n-1)/2} (r^2+b^2)^{n/2}} q.
\end{eqnarray}

It can be shown that $r=0$ is a curvature singularity. The event horizon $r_h$ is the largest root of equation $f(r_h)=0$. Depending on the free parameters $(\alpha, m, b)$ (or equivalently $(\alpha, m, q)$), there can be two horizons (inner and outer horizons), one degenerate horizon (extreme case) and naked singularity. The Hawking temperature on the event horizon is
\begin{eqnarray}\label{temperature}
T_H = \sqrt{r_h^2 + b^2} \frac{(N f)'}{4\pi \sqrt{N} r}\bigg|_{r=r_h}.
\end{eqnarray}
It is not possible to give an analytical expression of $r_h$. However, we can express the mass parameter $m$ in terms of $r_h$ by solving $f(r_h)=0$, which is
\begin{eqnarray}\label{massparameter}
m= \frac{(r_h^2 +b^2)^{n/2}}{\l^2} \Gamma_h^{(n-1)\gamma-1},
\end{eqnarray}
with $\Gamma_h = \Gamma(r=r_h)$. Substituting $m$ into Eq.~(\ref{temperature}), one obtains the temperature as
\begin{eqnarray}\label{temperature-1}
T_H = \frac{b\Gamma_h^{\gamma(n-1)/2-1}}{4\pi \l^2 (1-\Gamma_h)^{1/(n-2)}}\bigg\{(n-2)\big[(n-1)\gamma-1\big]-(n-1)\big[(n-2)\gamma-2\big]\Gamma_h\bigg\}.
\end{eqnarray}
From it, we can see that when $1>\gamma \geq \frac{1}{3} (\alpha \geq 1)$, the temperature is always positive $T_H>0$ and no extremal limit exists. While $0 < \gamma <\frac{1}{3} (\alpha<1)$, like a Reissner-Nordstr\"{o}m black hole, the temperature can be positive (non-extremal case) or zero (extremal case). In the latter case, if we think of all the parameters but $b$ fixed, then the maximum value of $b$ corresponding to the extremal case can be obtained by solving $T_H (b, r_h, \alpha)=0$, which is
\begin{eqnarray}\label{maxb}
b_{ext} = r_h \left\{\left(\frac{(n-1)[2-\gamma(n-2)]}{n}\right)^{2/(n-2)}-1\right\}^{-1/2}.
\end{eqnarray}
For more discussions on the thermodynamical properties of this black brane, see refs.~\cite{Sheykhi:2008rk,Sheykhi:2009pf,Hendi:2010gq,Ong:2012yf}.

According to AdS/CFT, the above black brane solution is dual to a CFT on the boundary with temperature given by Eq.~(\ref{temperature-1}). Moreover, the chemical potential $\mu$ of the boundary CFT is related to the asymptotical value of the temporal part of $A_\mu$, i.e., $\mu \sim \lim_{r\rightarrow \infty} A_t$. By integrating Eq.~(\ref{phi_F}), we can obtain the expression of $A_t$,
\begin{eqnarray}
A_t= -\frac{\sqrt{(n-1) m} b^{(n-2)/2}}{\sqrt{2(n-2+\alpha^2)} (r^2+b^2)^{(n-2)/2}} + \Psi_0,
\end{eqnarray}
Where $\Psi_0$ is a constant corresponding to the electrostatic potential at $r \rightarrow \infty$, which is defined such that the gauge field vanishes at the horizon, i.e.,
\begin{eqnarray}
\Psi_0 = \frac{\sqrt{(n-1) m} b^{(n-2)/2}}{\sqrt{2(n-2+\alpha^2)} (r_h^2+b^2)^{(n-2)/2}}.
\end{eqnarray}
The precise relation between the chemical potential and the gauge field is subtle because there is a dimension mismatch between the chemical potential in the dual field theory and the gauge field $A_\mu$ defined in the action Eq.~(\ref{action}). Thus, one has to rescale the gauge field as $\tilde{A}_\mu = \frac{A_\mu}{\xi}$, where $\xi$ is a scale with length unit that depends on the particular compactification. At this time, the chemical potential and the gauge field has the same unit and the duality relation is
\begin{eqnarray}
\mu = \lim_{r\rightarrow \infty} \tilde{A}_t = \frac{\Psi_0}{\xi}.
\end{eqnarray}
Since the boundary field theory we considered is conformal, the only relevant parameter we can vary is a dimensionless ratio $\chi$ constructed from the chemical potential and the temperature, i.e., $\frac{\mu}{T_H}$,
\begin{eqnarray}\label{chi}
\chi \equiv \frac{\mu}{T_H} = \frac{\Psi_0}{T_H \xi}.
\end{eqnarray}
Now we can see that if $\alpha$ and $m$ are kept fixed, then by varying $b$ which is equivalent to varying the charge parameter $q$, we can explore various values of the ratio $\chi$ in the dual field theory. For $0<\alpha<1$, when $b$ varies from $b=0$ (vanishing $\Psi_0$) to $b_{ext}$ (vanishing $T_H$), $\chi$ spans the whole range of values, i.e., from $\chi=0$ to $\infty$. However, for $\alpha\geq 1$, the situation becomes complicated. As we will see later, in this case, as $b$ runs over the whole range of values, $\chi$ is bounded in a finite range of values.

\subsection{Vaidya-like Solution}

Our goal is to investigate the thermalization process of the boundary field system under a quantum quench. According to AdS/CFT, this process can be holographically modeled by the process of a black hole formation in the bulk which can be described by a Vaidya-like metric. Thus, in this section, we would like to get the Vaiya-like metric by generalizing the black brane solution to be time-dependent.

It is more convenient to work with a new radial coordinate $z \equiv \frac{\l^2}{r}$ such that the AdS boundary lies at $z=0$. Thus, the metric becomes
\begin{eqnarray}\label{metric-1}
ds^2 &=& -N(z) f(z) dt^2 + \frac{\l^8}{z^4} \frac{dz^2}{(l^4+b^2 z^2) f(z)} + \frac{\l^4+b^2 z^2}{z^2} g(z) d\vec{x}^2_{n-1}, \nonumber\\
N(z) &=& \Gamma^{-(n-3)\gamma},\qquad f(z) = \frac{\l^4+b^2 z^2}{\l^2 z^2} \Gamma^{(n-2)\gamma}-m \left(\frac{z}{\sqrt{\l^4+b^2 z^2}}\right)^{n-2} \Gamma^{1-\gamma},\nonumber\\
g(z) &=& \Gamma^\gamma,\qquad\qquad\quad~~~ \Gamma = 1-\left(\frac{b z}{\sqrt{\l^4 +b^2 z^2}}\right)^{n-2}.
\end{eqnarray}
Introducing ingoing Eddington-Finkelstein coordinate
\begin{eqnarray}\label{v}
d v = dt - \frac{\l^4}{z^2} \frac{d z}{\sqrt{(\l^4+b^2 z^2) N(z)} f(z)},
\end{eqnarray}
the above metric becomes
\begin{eqnarray}\label{metric-2}
ds^2 &=& - N(z) f(z) dv^2 - \frac{2 \l^4}{z^2} \sqrt{\frac{N (z)}{\l^4+b^2 z^2}} d v d z+ \frac{l^4+b^2 z^2}{z^2} g(z) d\vec{x}^2_{n-1},\nonumber\\
f(z) &=& \frac{\l^4+b^2 z^2}{\l^2 z^2} \Gamma^{(n-2)\gamma}-m \left(\frac{z}{\sqrt{\l^4+b^2 z^2}}\right)^{n-2} \Gamma^{1-\gamma}.
\end{eqnarray}
Usually, the Vaidya-like version can be obtained by using time dependent  mass and charge parameters $(m\to m(v), q\to q(v))$, that is,
\begin{eqnarray}
m(v) = m T(v),\qquad q(v)=q T(v)^{1/2},
\end{eqnarray}
where $m$ and $q$ are the values of the corresponding parameters at late times. We can still introduce the parameter $b$, which satisfies the relation Eq.~(\ref{chargeparameter}), to make our expression explicitly. After taking such time-dependent parameters, we obtain the Vaiya-like metric which takes the same form as Eq.~(\ref{metric-2}) but with $f(z)$ being now
\begin{eqnarray}
f(v, z) &=& \frac{\l^4+b^2 z^2}{\l^2 z^2} \Gamma^{(n-2)\gamma}-m T[v] \left(\frac{z}{\sqrt{\l^4+b^2 z^2}}\right)^{n-2} \Gamma^{1-\gamma}.
\end{eqnarray}
Now, the equations of motion Eqs.~(\ref{EoM}) do not hold any more and external matter sources need to be introduced,
\begin{eqnarray}
R_{\mu\nu}-\frac{1}{2} g_{\mu\nu} R- \frac{4}{n-1} \left(\partial_\mu \Phi \partial_\nu \Phi-\frac{1}{2} g_{\mu\nu} \left(\nabla \Phi\right)^2-\frac{n-1}{8} g_{\mu\nu} V(\phi)\right)\nonumber\\
-2 e^{-4\alpha \Phi/(n-1)} \left(F_{\mu\rho} F_{\nu}^{~\rho}-\frac{1}{4} g_{\mu\nu} F_{\rho\sigma} F^{\rho\sigma}\right)&=& 8\pi T_{\mu\nu}^{(ext)},\nonumber\\
\nabla_\mu \left(e^{-4\alpha \Phi/(n-1)} F^{\mu\nu}\right)&=&8\pi J^\nu_{ext}.
\end{eqnarray}
By checking the equation of motion of the dilaton field, we find that it always holds and no more additional external terms need to be introduced.

In order to keep simplicity, in this paper we only consider the case with $n=4$, that is, the dual field theory we considered is a four-dimensional conformal field theory. The above Vaidya-like metric is a solution of the equations of motion Eqs.~(\ref{EoM}) provided the external matter source satisfies
\begin{eqnarray}
8\pi T_{\mu\nu}^{(ext)} &=& \left\{\frac{3 z^3 \left[(2+\alpha^2) \l^4 + \alpha^2 b^2 z^2\right]}{2(2+\alpha^2) (\l^4 + b^2 z^2)^{5/2}} \left(\frac{\l^4}{\l^4+ b^2 z^2}\right)^{-\frac{3\alpha^2}{2(2+\alpha^2)}} m \dot{T} (v)\right\} \delta_\mu^v \delta_\nu^v,\nonumber\\
8\pi J^\nu_{(ext)} &=& \left\{\frac{\sqrt{3} b z^5}{\l^8 \sqrt{2(2+\alpha^2)}} \left(\frac{\l^4}{\l^4+b^2 z^2}\right)^{\frac{2}{2+\alpha^2}}\sqrt{m} \frac{\dot{T}(v)}{\sqrt{T(v)}}\right\} \delta_z^\nu,
\end{eqnarray}
where dot means $\partial_v$. We can see that the in-falling matter is charged dust. As  in refs.~\cite{Balasubramanian:2010ce,Balasubramanian:2011ur}, we consider $T(v)$ to take the form
\begin{eqnarray}
T(v) &=& \frac{1}{2} \left(1 + \tanh\frac{v}{v_0}\right),
\end{eqnarray}
where $v_0$ is a small parameter, with typical order of $10^{-3}$. That is, the ingoing falling matter is a thin shell. Thus, in the bulk gravity side, we consider the collapsing of an ingoing falling thin-shell to form a black hole. And this picture is dual to a sudden energy injection into the field system  evolving  towards thermal equilibrium.

\section{Non-local Observables}

Now, we will use the Vaidya-like metric constructed in last section to discuss the thermalization process of the dual field system. To investigate the details of the thermalization, one can use three non-local observables, the two-point coorelation function, the Wilson loop and the entanglement entropy, to probe the process. According to AdS/CFT, in the saddle approximation these three non-local observable can be evaluated by calculating the corresponding geometric quantities in the bulk geometry.

As we stated above, we mainly focus on the case with $n=4$. In doing numerical calculations in the following, we would like to set $\l=m=1$. That is, we fix the mass of the black brane formed finally, which is equivalent to fix the amount of energy injected into the dual boundary field system. And then we would like to see the effect of the chemical potential (more precisely $\chi$) and the coupling constant $\alpha$ on the thermalization process.  From Eq.~(\ref{massparameter}), we can see that, after the mass parameter $m$ is fixed, the event horizon $r_h$ is determined by parameters $\alpha$ and $b$ through the relation
\begin{eqnarray}
(r_h^2+b^2)^2 \Gamma_h^{3\gamma-1}=1,
\end{eqnarray}
with $\Gamma_h = \frac{r_h^2}{r_h^2+b^2}$. When $\alpha> 1$, there is no bound of $b$ and it can take arbitrary values. When $\alpha=1$, the above relation reduces to $r_h^2 + b^2 =1$, so $b$ should be less than $1$ to guarantee the existence of the horizon. While $0<\alpha<1$, from Eq.~(\ref{maxb}) the maximum value of $b$ is
\begin{eqnarray}
b_{ext} = r_h \sqrt{\frac{2}{1-3 \gamma}}.
\end{eqnarray}
The charge parameter $q = \sqrt{\frac{6}{2+\alpha^2}} b$ then takes values in the range $q \in [0,q_{ext}]$ at this time, with $q_{ext}=\sqrt{\frac{6}{2+\alpha^2}} b_{ext}$.

From Eq.~(\ref{chi}), the ratio of the chemical potential over temperature $\chi$ is
\begin{eqnarray}
\chi \equiv \frac{\mu}{T_H} = \frac{4\pi \sqrt{3} b}{\sqrt{2(2+\alpha^2)} (r_h^2+b^2)^{3/2} \Gamma_h^{3\gamma/2 - 1}} \left[2(3\gamma-1)-3(2\gamma-2) \Gamma_h\right]^{-1},
\end{eqnarray}
where we have taken the scale $\xi=1$. The relation between $\chi$ and $(q, \alpha)$ is plotted in Fig. 1, from where we can see that fixing $\alpha$, $0<\alpha < 1$ and increasing $q$  corresponds to increasing $\chi$. When $q$ varies from $q=0$ to $q=q_{ext}$, the value of $\chi$ varies from $0$ to $\infty$. When $\alpha > 1$, $\chi$ is no longer a monotonically increasing function of $q$. In fact, as $q$ increases from $0$ to $\infty$, $\chi$ first increases and after reaching a maximum value at $q_{max}$ it begins to decrease. In this case, the value of $\chi$ is bounded in a finite range. For $\alpha=1$, the charge $q$ should be less than $\sqrt{2}$ to guarantee the existence of the horizon as we have stated above. From the figure, we can see that in this case, $\chi$ is still a monotonically increasing function of $q$ and approaches a finite constant as $q$ approaches $\sqrt{2}$.

\begin{figure}[!htbp]
\centering
\includegraphics[width=0.5\textwidth]{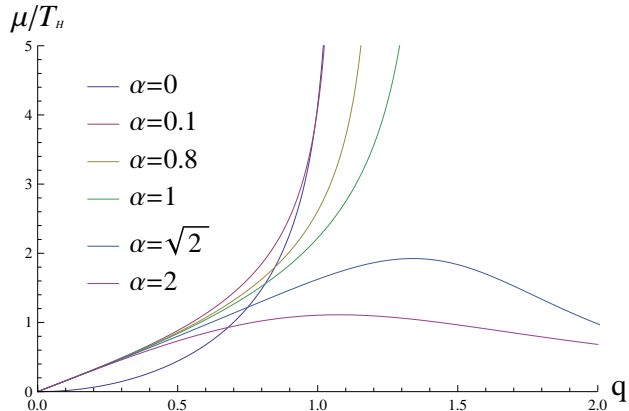}
\caption{The chemical potential over the temperature versus $q$ in $n=4$ case for various $\alpha$. For $\alpha=1$, $q$ should be less than $\sqrt{2}$ to guarantee the existence of the horizon. When $\alpha>1$, there is a maximum point of $\mu/T_H$ at $q_{max}$, which is $q_{max}=\sqrt{6}/(5^{3/8}) \approx 1.3395$ for $\alpha=\sqrt{2}$ and  $q_{max}=\sqrt{\frac{2}{\sqrt{3}}} \approx 1.0745$ for $\alpha=2$.}
\end{figure}

\subsection{The two-point fuction}

On the boundary at time $t$, we choose two points with spatial separation $L$ and compute the correlation function of some operator with large conformal dimension located at the two points. Considering the symmetry of the boundary, we can always choose coordinates such that the two points lie in the $x_1$-axis with coordinates $(t, x_1)=(t, -\frac{L}{2}), (t, \frac{L}{2})$ respectively. Other coordinates of the two points are identical. According to the AdS/CFT duality, in the saddle point approximation the two-point function can be evaluated by calculating the length of the space-like geodesic connecting the two points. The geodesic can be parameterized by only two functions: $v(x)$ and $z(x)$, with other coordinates unchanged, where we simply denote $x_1 \equiv x$. Thus, the induced metric on the geodesic is
\begin{eqnarray}
ds^2 = \left(-N (z) f (v, z) v'^2-\frac{2}{z^2} \sqrt{\frac{N(z)}{1+b^2 z^2}} v' z' + \frac{1+b^2 z^2}{z^2} g(z)\right) dx^2.
\end{eqnarray}
The length functional of the geodesic is
\begin{eqnarray}\label{lengthfunctional}
{\cal L} &=& 2\int_0^{L/2} P^{1/2} dx,\\
P &\equiv& -N (z) f(v, z) v'^2-\frac{2}{z^2} \sqrt{\frac{N(z)}{1+b^2 z^2}} v' z' + \frac{1+b^2 z^2}{z^2} g(z).\nonumber
\end{eqnarray}
To minimize the length of the geodesic, we need to solve the two equations of motion for $v(x)$ and $z(x)$ respectively. The expressions are rather involved and we do not give them explicitly here. Considering the symmetry, we have in mind the picture of the geodesic: starting from one of the two points on the boundary $z=0$, the geodesic extends into the bulk and after reaching the turning point $(v(0), z(0))=(v_\ast, z_\ast)$ it turns back and finally ends at another point on the boundary. At the turning point $x=0$, the  derivatives vanish, that is, $v'(0) = z'(0) = 0$. It seems that one can impose the boundary conditions at $x=0$ to solve the two equations of motion. However, the two equations of motion are singular at $x=0$, and in practice, to make our numerical calculations available, we would rather impose the boundary conditions at the neighborhood of the turning point $x=0$,
\begin{eqnarray}\label{bnycondition}
v(\epsilon) = v_\ast + {\cal O}(\epsilon^2),\quad v'(\epsilon)={\cal O} (\epsilon^2), \quad z(\epsilon) = z_\ast + {\cal O}(\epsilon^2), \quad z'(\epsilon)={\cal O} (\epsilon^2),
\end{eqnarray}
where $\epsilon$ is a small quantity, with typical order of $10^{-3}$. The ${\cal O} (\epsilon^2)$ terms are corrections terms and can be derived by solving the two equations of motion around $x=0$ order by order. In our calculations, we only calculate to order $\epsilon^2$. The two free parameters $v_\ast$ and $z_\ast$ characterizing the turning point of the geodesics are determined by the constraint equations
\begin{eqnarray}
v (\pm L/2) = t,\qquad z(\pm L/2) =z_0,
\end{eqnarray}
where $z_0$ is a UV-cutoff to make the length of the geodesic finite and $t$ is the boundary time. Furthermore, note that the integrand in the length functional Eq.~(\ref{lengthfunctional}) does not depend on $x$ explicitly. So if we treat $x$ as a "time" variable then the corresponding Hamiltonian is conserved, which yields the following equation
\begin{eqnarray}\label{conservation}
P = \left(\frac{1+b^2 z_\ast^2}{z_\ast^2} g(z_\ast)\right)^{-1} \left(\frac{1+b^2 z^2}{z^2} g(z)\right)^2.
\end{eqnarray}
By using it, the length functional can be rewritten as
\begin{eqnarray}
{\cal L} = 2\int_0^{L/2} dx \left(\frac{1+b^2 z_\ast^2}{z_\ast^2} g(z_\ast)\right)^{-1/2} \left(\frac{1+b^2 z^2}{z^2} g(z)\right).
\end{eqnarray}

The length of the geodesic ${\cal L}$ is a function of the boundary time $t$. By calculating the length at different boundary times, we can get the time evolution of the two-point function. For convenience, we define a rescaled length $\delta \tilde{\cal L} \equiv \frac{{\cal L}-{\cal L}_{thermal}}{L}$ and study its time evolution. To make the length independent of the cutoff $z_0$, we have subtracted the length at late time in the definition. $\delta \bar{\cal L}$ depends on three parameters: the boundary time $t$, the dilaton-Maxwell coupling constant $\alpha$ and the charge $q$. By varying $\alpha$ and $q$, we can study the effect of the two parameters on the time evolution of the two-point function.

\begin{figure}[!htbp]
\centering
\subfigure[$~~\alpha=0$]{\includegraphics[width=0.4\textwidth]{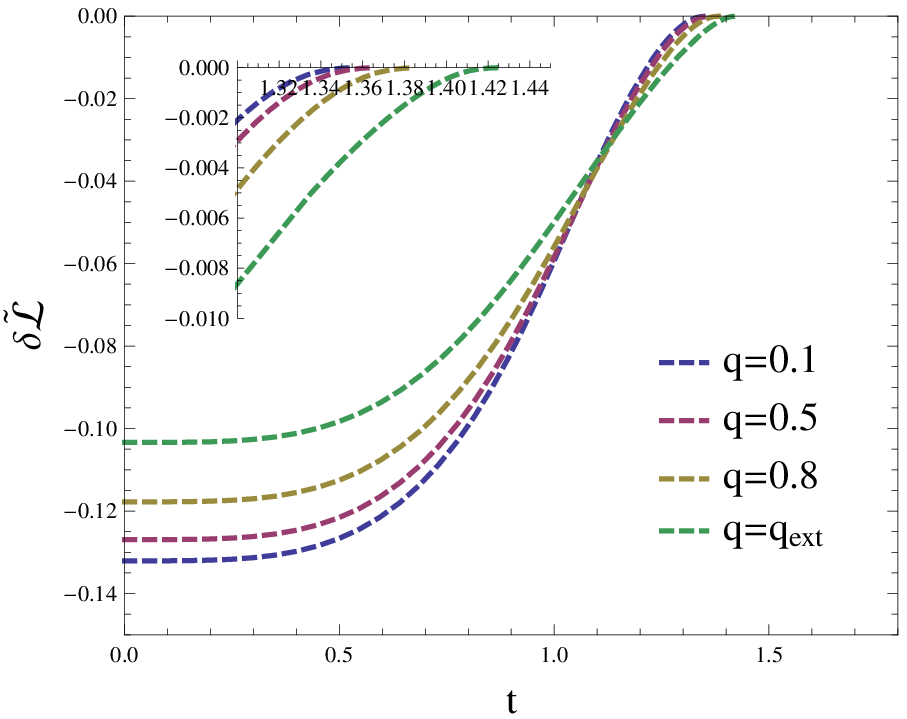}}\quad
\subfigure[$~~\alpha=0.1$]{\includegraphics[width=0.4\textwidth]{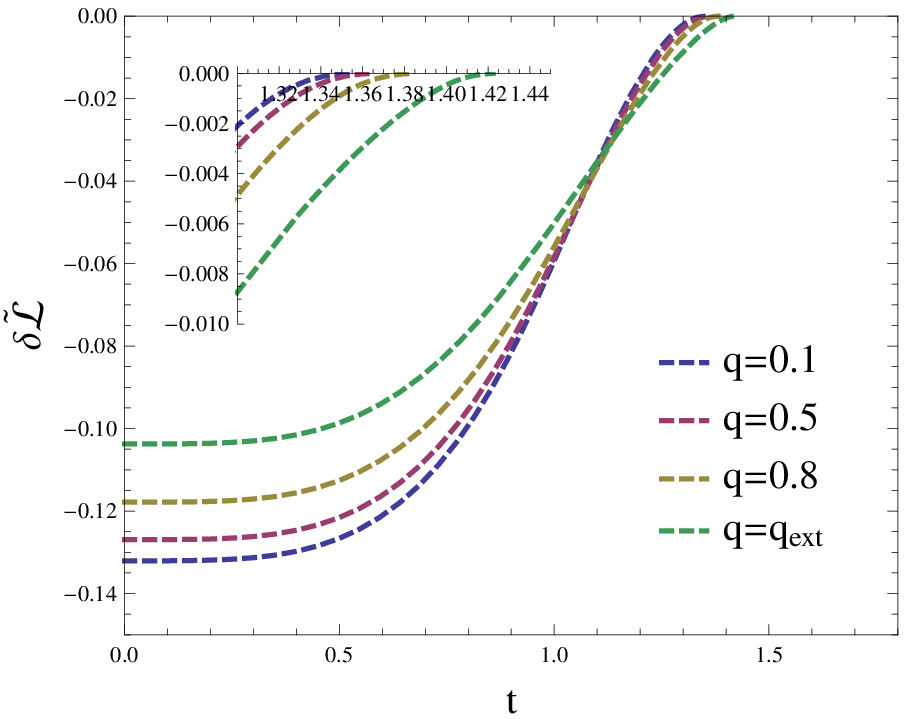}}
\subfigure[$~~\alpha=0.8$]{\includegraphics[width=0.4\textwidth]{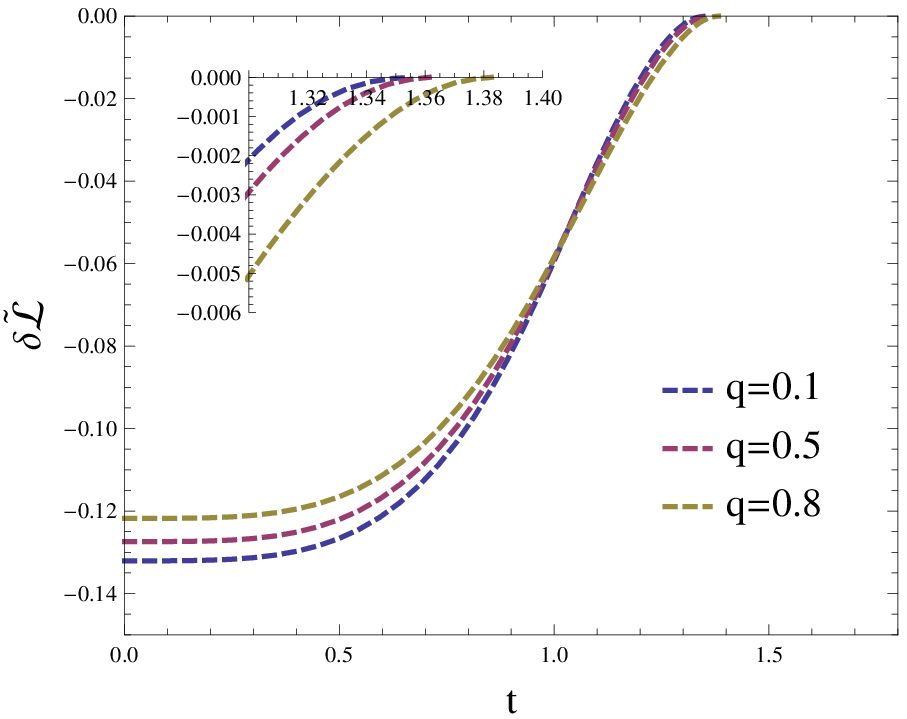}}\quad
\subfigure[$~~\alpha=1$]{\includegraphics[width=0.4\textwidth]{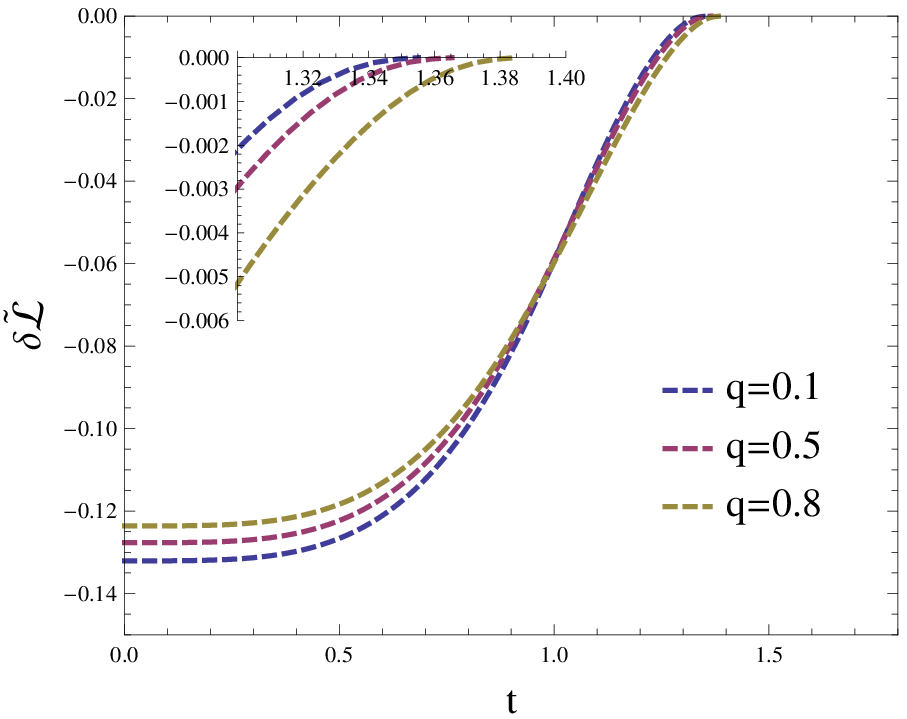}}
\caption{Time evolution of the two-point function for various values of $\alpha$ and $q$. Insets in each graph zoom in the late time behavior of these curves. The spatial separation of the two points are $L=3$. For $\alpha =0$, $q_{ext} = \sqrt{\frac{2}{\sqrt{3}}} \approx 1.0745$; For $\alpha =0.1$, $q_{ext} \approx 1.0763$; For $\alpha =0.8$, $q_{ext} \approx 1.22386$; For $\alpha=1$, $q$ should be smaller than $\sqrt{2}$ to guarantee the existence of the horizon.}
\end{figure}

In Fig. 2, we plot the time evolution of the two-point function for various values of $\alpha$ $(0 \leq \alpha \leq 1)$ and $q$. For comparison, we also give the result for the $\alpha=0$ case. The spatial separation of the two points on the boundary is chosen to be $L=3$. From the figure, we can see that that the whole process can be divided in the following five successive stages: delay at early times, pre-local-equilibrium with  quadratic growth in time, post-local-equilibrium with linear growth, memory-loss-period, and  equilibrium. These features have been observed in various holographic models for thermalization and can be considered to be universal for strongly coupled field systems.

As we increase the charge $q$,  equivalent to increasing the ratio of the chemical potential over temperature $\chi$ of the dual field system for $0 \leq \alpha \leq 1$, the initial absolute value of $\delta \tilde{{\cal L}}$ decreases, meaning that the dual field system is initially closer to thermal equilibrium. However, larger $q$ makes the delay time longer and the growth of $\delta \tilde{{\cal L}}$ slower, thus the saturation time to reach thermal equilibrium increases. This means, for $0 \leq \alpha \leq 1$, when we fix the energy injected, that the saturation time increases as $\chi$ increases. Also, we should emphasize that the dependence of the saturation time on $\chi$ is weak, as we can see from the figure. This dependence of the saturation time on the ratio of the chemical potential over temperature $\chi$ has been observed in the case with standard Maxwell term~\cite{Galante:2012pv,Caceres:2012em,Zeng:2013fsa,Giordano:2014kya} and also with non-linear Born-Infeld term~\cite{Camilo:2014}.

However, when $\alpha>1$, the situation becomes complicated, as we show in Fig. 3 with $\alpha=2$. In this case, the initial absolute value of $\delta \tilde{{\cal L}}$ is no longer a monotonically decreasing function of the charge $q$. In fact, as $q$ increases, the initial absolute value of $\delta \tilde{{\cal L}}$ first decreases and then after $q$ crosses some critical value,  the initial absolute value of $\delta \tilde{{\cal L}}$ begins to increase. More interesting, we can see from the figure that the saturation time is still a monotonically increasing function of $q$. However, from Fig. 1, we can see that for $\alpha>1$, $\chi$ is no longer a monotonically increasing function of $q$, thus the saturation time is no longer a monotonically increasing function of $\chi$: When $q \leq q_{max}$, larger $\chi$ yields longer $t_{sat}$; However, when $q>q_{max}$, we have the inverse behavior. We can see it more clearly in the right panel of Fig. 3.
\begin{figure}[!htbp]
\centering
\includegraphics[width=0.4\textwidth]{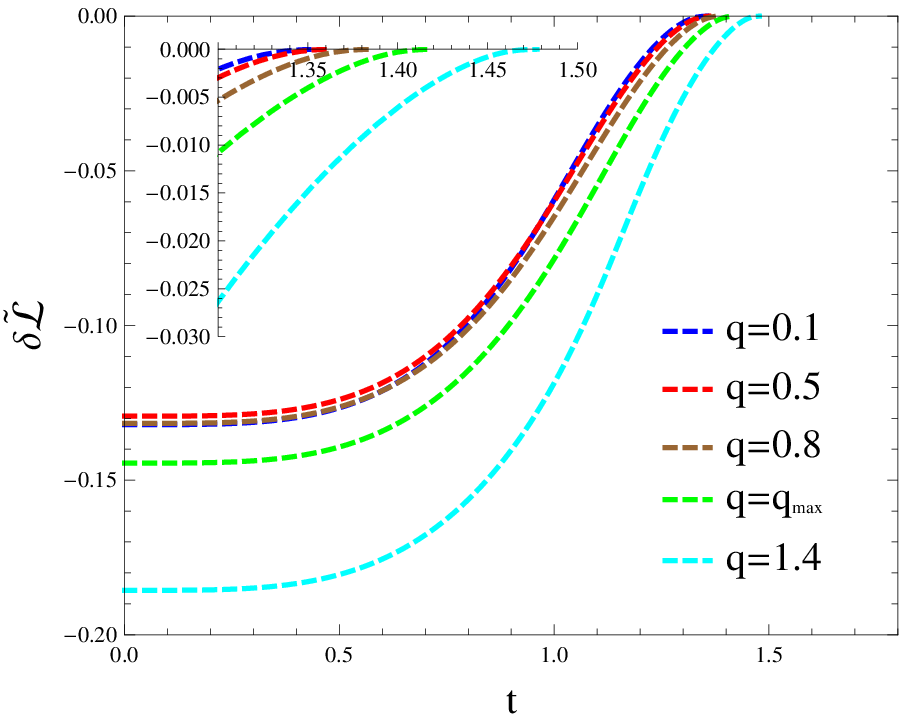}\qquad\qquad
\includegraphics[width=0.4\textwidth]{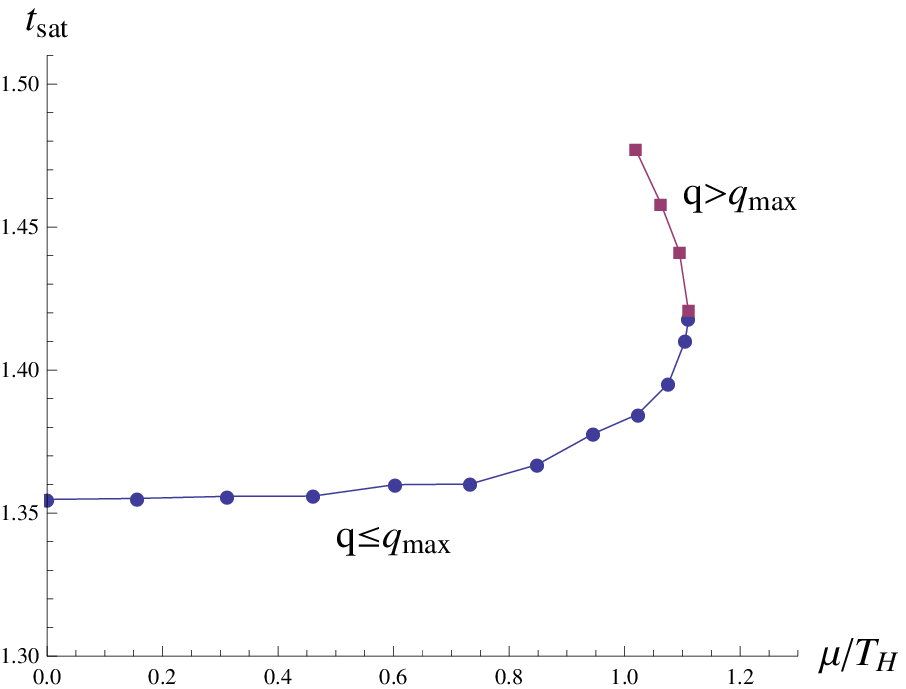}
\caption{{\em Left:} Time evolution of the two-point function in $\alpha=2$ case with $L=3$. There is no bound for $q$ in this case. $q_{max} \approx 1.0745$ corresponds to the point when $\frac{\mu}{T_H}$ reaches its maximum value. There is a large overlap between the curves with $q=0.1$ and the one with $q=0.8$ at early times. {\em Right:} Saturation time $t_{sat}$ versus the ratio of chemical potential over temperature $\chi$ for $\alpha=2$ case.}
\end{figure}

By comparing the four graphs with different values of $\alpha$ in Figs. 2 and 3, one can see that the effect of the coupling constant $\alpha$ on the saturation time is nearly negligible. We can see this point more clearly from the left panel of Fig. 4, where we plot the time evolution of the two-point function for various values of $\alpha$ and fixed $q$. From the figure, we can see that, for fixed $q$, larger $\alpha$ makes the initial state of the dual field system further away from thermal equlibrium, but will make the delay time shorter and the growth of $\delta \tilde{{\cal L}}$ faster thus yielding the saturation time nearly un-affected. The saturation time is almost independent of the coupling constant $\alpha$. As the one-parameter gravity action Eq.~(\ref{action}) corresponds to a one-parameter dual field theory, this phenomenon indicates the universality of the saturation time of this one-parameter strongly coupled field system to reach thermal equilibrium under quantum quench.

\begin{figure}[!htbp]
\centering
\includegraphics[width=0.42\textwidth]{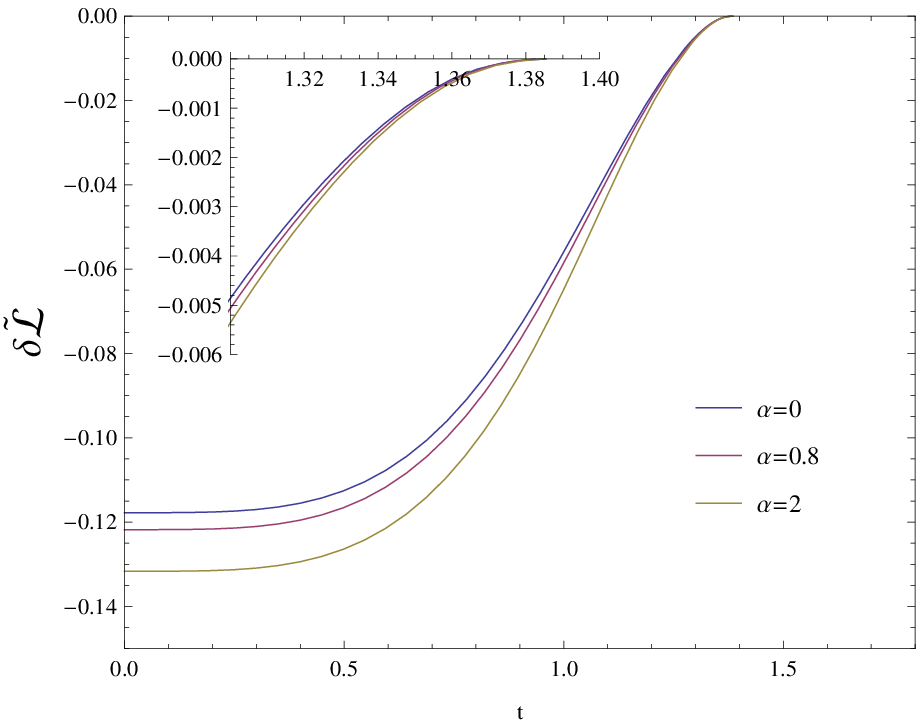}\qquad\qquad
\includegraphics[width=0.4\textwidth]{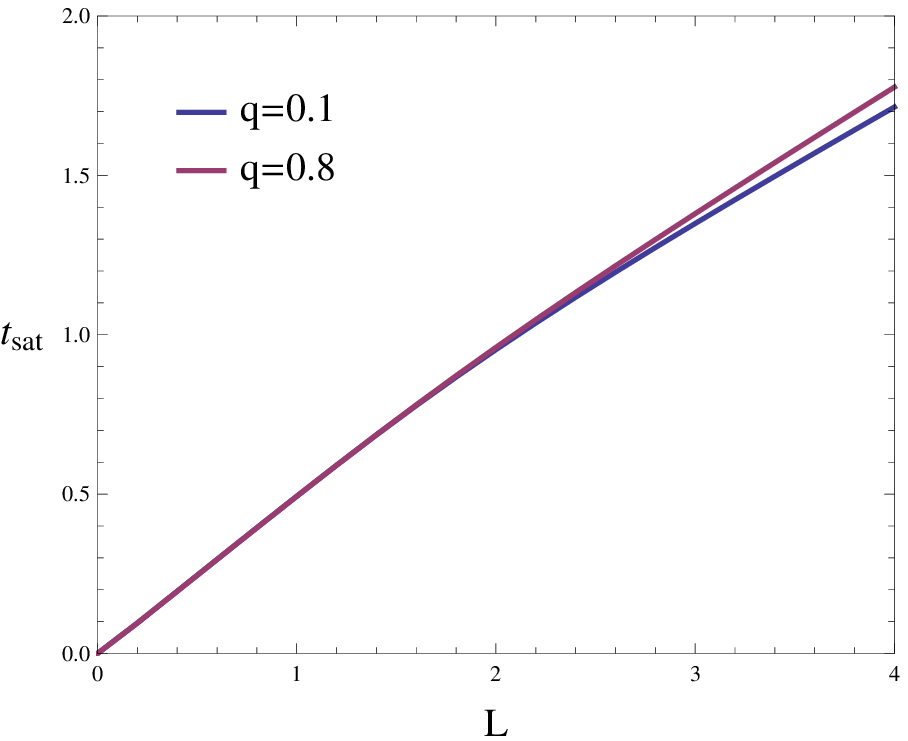}
\caption{{\em Left}: Time evolution of the two-point function for various $\alpha$ with $q=0.8$. The spatial separation of the two points is $L=3$. {\em Right}: Saturation time versus the boundary scale. The result is the same for various values of $\alpha$.}
\end{figure}

The saturation time depends on the boundary scale $L$, as we can see from the right panel of Fig. 4, where we plot the saturation time versus $L$. Note that the effect of $\alpha$ on the saturation time can be neglected as we stated above, thus this figure is correct for various $\alpha$. As we can see from the figure, the saturation time is linear in $L$  with slope less than unit when $L$ is small, and sub-linear when $L$ becomes large. Once again, we can see that the saturation time depends weakly on the charge: larger $q$ will have longer saturation time.

\subsection{The Wilson loop and entanglement entropy}

According to the AdS/CFT duality, the expectation value of the Wilson loop of the boundary field theory, in the saddle approximation, is dual to the area of a two-dimensional extremal surface in the bulk which anchored on the loop at the boundary. We choose the Wilson loop to be a circle and study its time evolution during thermalization. It turns out that the time evolution of this observable shows a similar behavior as that of the two-point function, which supports our results above. So to avoid redundancy, we will not show the results for the Wilson loop here.

Now we turn to use the entanglement entropy to probe the thermalization process. Among the three non-local observables, it involves  most degrees of freedom of the boundary field system. So we expect that by studying its evoluton, we can gain more information of the process. On the boundary, we choose a spherical subregion of the spatial space constrained by the relation $\sum_{i=1}^3 x^2_{i-1} \leq R^2$, and calculate its entanglement entropy. According to the conjecture proposed by refs.~\cite{Ryu:2006bv,Hubeny:2007xt}, the entanglement entropy of a subregion of the boundary spatial space is dual to the area of a codimension-two extremal surface anchored on the boundary of the chosen subregion. Using spherical coordinates $(\varrho,\Omega_2)$ and considering the spherical symmetry, the codimension-two extremal surface can be parameterized by only two functions, $v(\varrho)$ and $z(\varrho)$. Thus, the induced metric on the surface is
\begin{eqnarray}
ds^2 = \left(-N (z) f(v, z) v'^2-\frac{2}{z^2} \sqrt{\frac{N(z)}{1+b^2 z^2}} v' z' + \frac{1+b^2 z^2}{z^2} g(z)\right) d\varrho^2 + \frac{1+b^2 z^2}{z^2} g(z) \varrho^2 d\Omega_2^2,
\end{eqnarray}
and the area functional is
\begin{eqnarray}\label{EEfuntional}
S &=& 4 \pi \int_0^{R} \frac{\varrho^2 g(z) \sqrt{1+b^2 z^2}}{z^2} P^{1/2} d\varrho,\\
P &\equiv& -N (z) f(v, z) v'^2-\frac{2}{z^2} \sqrt{\frac{N(z)}{1+b^2 z^2}} v' z' + \frac{1+b^2 z^2}{z^2} g(z).\nonumber
\end{eqnarray}
To get the extremal value of the functional, as done in previous two subsections, we need to solve the two equations of motion, not shown here.  The same boundary conditions as Eq.~(\ref{bnycondition}) are imposed. We define a rescaled area $\delta \tilde{{\cal S}} \equiv \frac{{\cal S}-{\cal S}_{thermal}}{\frac{4}{3} \pi R^3}$, where $\frac{4}{3} \pi R^3$ is the volume of the chosen subregion, and study its time evolution during thermalization.

\begin{figure}[!htbp]
\centering
\subfigure[$~~\alpha=0$]{\includegraphics[width=0.4\textwidth]{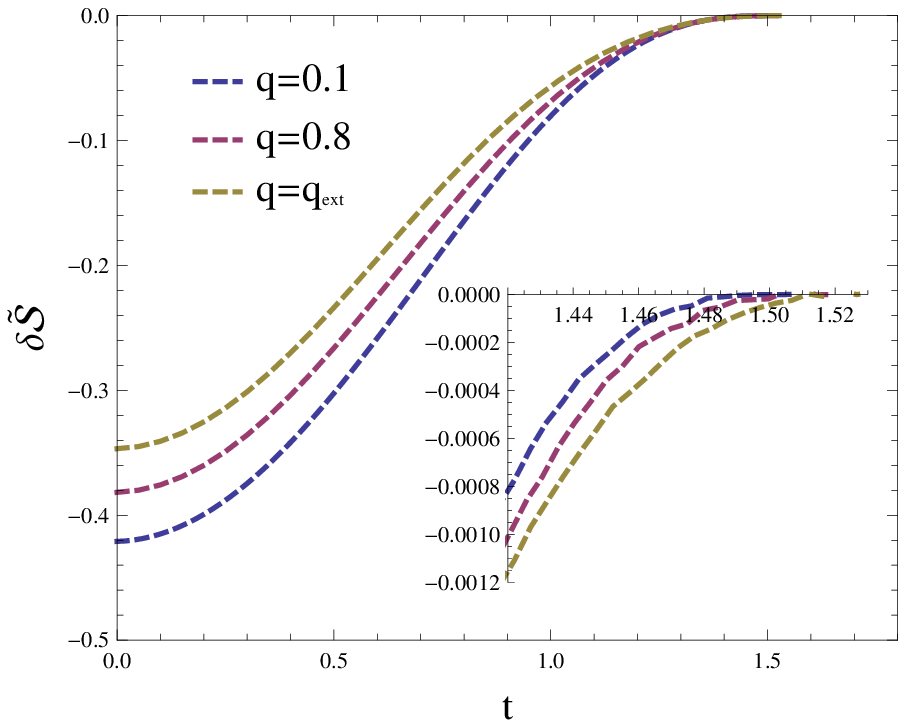}}\quad
\subfigure[$~~\alpha=0.1$]{\includegraphics[width=0.4\textwidth]{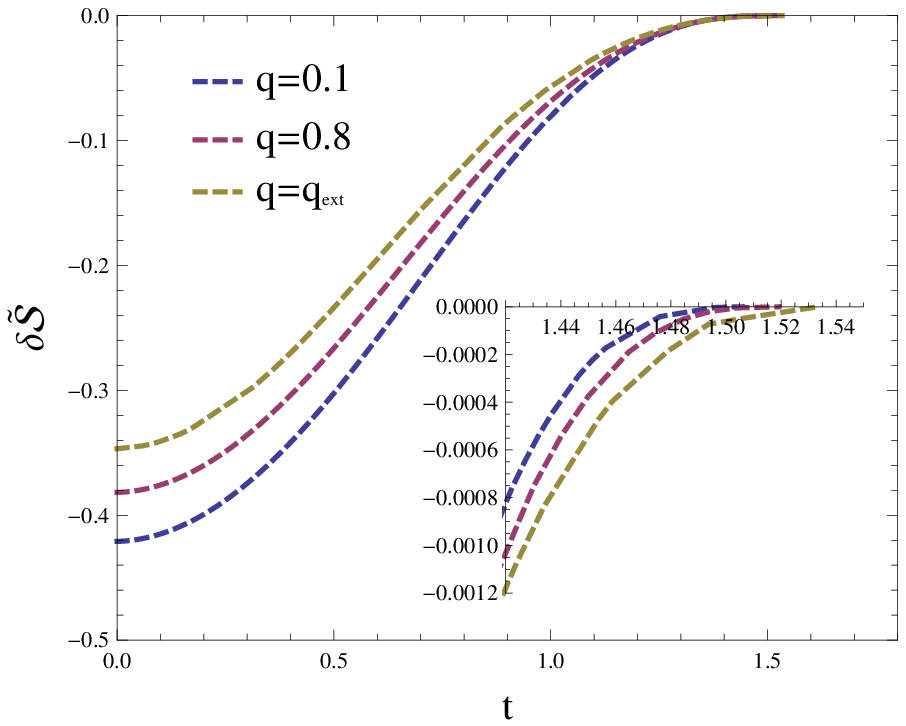}}
\subfigure[$~~\alpha=0.8$]{\includegraphics[width=0.4\textwidth]{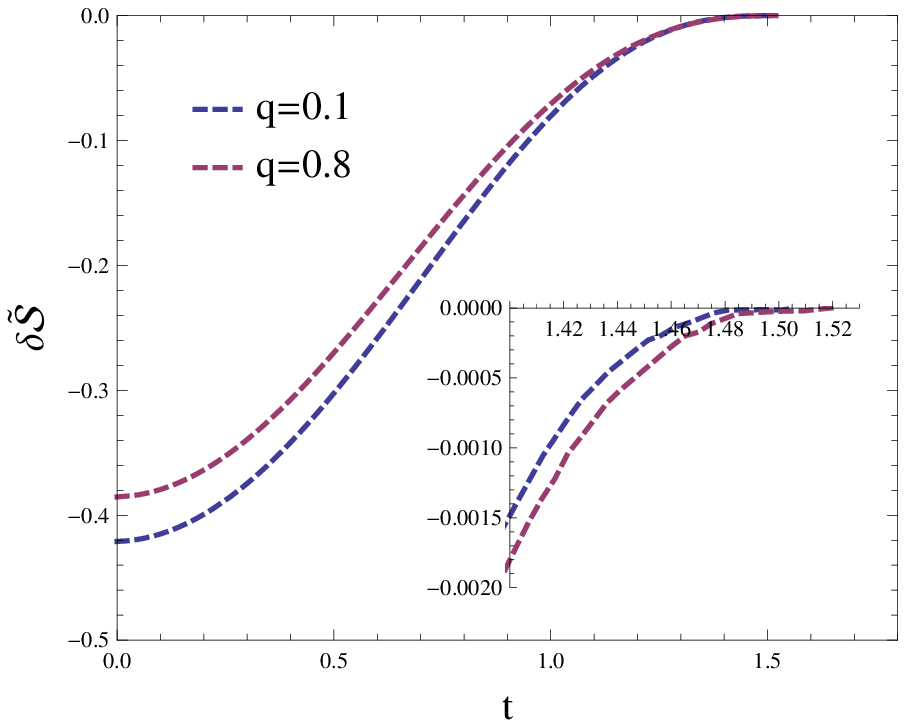}}\quad
\subfigure[$~~\alpha=2$]{\includegraphics[width=0.4\textwidth]{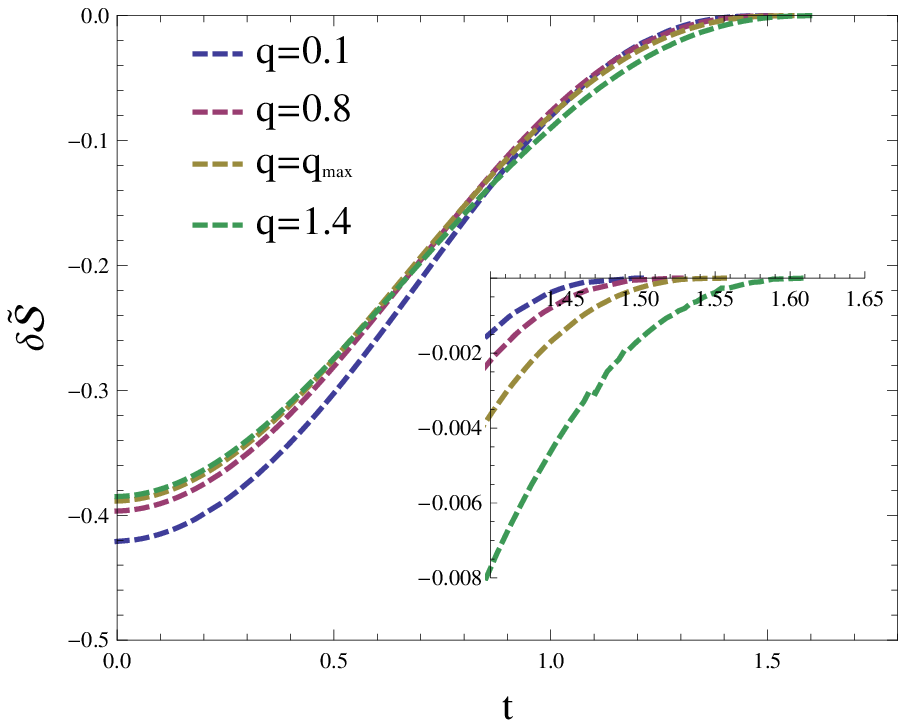}}
\caption{Time evolution of the entanglement entropy for various $\alpha$ and $q$. The inset in each graph zooms in the late time behavior of each curve. The radius of the entanglement sphere is $R=1.5$.}
\end{figure}

In Fig. 6, we plot the time evolution of the entanglement entropy for various values of $\alpha$ and $q$. The evolution shows a similar behavior as those of the other two non-local observables: for fixed $\alpha$, the saturation time increases as the charge $q$ increases. For $0 \leq \alpha \leq 1$, this means the ratio of the chemical potential over temperature $\chi$ enhances the saturation time. While for $\alpha>1$, the situation depends on the charge $q$: when $q \leq q_{max}$, the effect of $\chi$ on the saturation time is to enhance, while the effect is opposite when $q>q_{max}$. Moreover, we see once again that the dilaton-Maxwell coupling constant $\alpha$ has almost no influence on the saturation time, although it indeed affects the whole process. These results provide further support for the ones obtained from the other two observables.

However, there are some differences comparing to the results of other two non-local observables. First, the delay time for the entanglement entropy is the shortest. This is understandable, as entanglement entropy, among the three non-local observables involves most degrees of freedom of the field system and thus is the most sensitive to thermalization. Second, now for all cases with different $\alpha$ we show, the behavior of the initial absolute value of $\delta \tilde{\cal L}$ as $q$ increasing is the same: it decreases as $q$ increases. This is different from results of the other two observalbes, where the initial absolute value of $\delta \tilde{\cal L}$ is not a monotonically decreasing function of $q$ when $\alpha >1$. This is understandable, as the three observables relate to different amounts of degrees of freedom of the boundary field system, and thus may reflect different aspects of the system. As we stated above, the entanglement entropy involves more degrees of freedom and thus can can reflect more clearly the average state of the field system. Thus it is expected that if all the degrees of freedom of the field system are taken into account, the initial state of the field system gets closer to the thermal equilibrium as $q$ increases.

\section{Summary and Discussions}

We studied the holographic thermalization process in the presence of a chemical potential and dilaton field. Three non-local observables are used to probe the process, and they together show a rich and consistent picture of the thermalization. The effect of the chemical potential (or the charge) and the dilaton field on the thermalization is explored in detail. The saturation time increases as the charge $q$ increases, which means that the ratio of the chemical potential over temperature $\chi$ enhances the saturation time when $0 \leq \alpha \leq 1$. However, when $\alpha$ is larger than $1$, the situation becomes a little complicated: when $q$ is smaller than some critical value $q_{max}$, the saturation time increases as the $\chi$ increases; however, when $q>q_{max}$, the behavior is inverse. This is due to the fact that the $\chi$ is no longer a monotonically increasing function of $q$ for $\alpha>1$. This is different from cases with standard Maxwell term~\cite{Galante:2012pv,Caceres:2012em,Zeng:2013fsa} or non-linear Born-Infeld term~\cite{Camilo:2014}, where $\chi$ is always a monotonically increasing function of $q$.

We also investigate the effect of the dilaton-electromagnetic coupling constant $\alpha$ on the thermalization process. This non-trival coupling yields great effects on the availabe background spactimes. However, it is remarkable to see that it nearly has no influence on the saturation time. As different $\alpha$ in the action defines a different gravity theory and thus corresponds to a different dual field theory, it indicates the universality of the saturation time of this one-parameter field theories to reach thermal equilibrium under quantum quench.  Although it has no effect on the saturation time, it indeed leave imprints on the whole thermalization process. If we fixed the charge $q$, increasing $\alpha$ makes the initial state of the dual field system further away from thermal equlibrium, the delay time shorter and the growth of $\delta \tilde{{\cal L}}$ faster. It is interesting to see if an physical explanation on this point can be given.

We have modelled the thermalization process via the naive gravitational collapse of a charged thin shell. It can also be modelled by the collapsing of a scalar field as done in refs.~~\cite{Danielsson:1999fa,Janik:2006gp,Chesler:2009cy,Garfinkle:2011hm,Garfinkle:2011tc}. It is interesting to see if this more realistic model can give us more information about the thermalization process. We leave it for further investigations.

\section*{Acknowledgments}

This work has been supported by FAPESP and CNPq (Brazil).

\end{document}